\documentclass[conference,final,twocolumn,10pt]{IEEEtran} 

\ifCLASSINFOpdf
  \usepackage[pdftex]{graphicx}
  \usepackage{epstopdf} 
\else
\fi
\usepackage{amsmath}
\usepackage{algorithm}
\usepackage{algorithmic}

  \usepackage[caption=false,font=footnotesize]{subfig}
\usepackage{url}

\usepackage{amsthm}

\theoremstyle{definition}

\theoremstyle{remark}

\newcommand{\comment}[1]{{}}

\usepackage{amssymb}
\usepackage{xspace}
\usepackage{bm}

\newcommand{\mat}[1]{\ensuremath{\mathbf{#1}}\xspace} 
\renewcommand{\vec}[1]{\ensuremath{\mathbf{#1}}\xspace} 

\newcommand{\parens}[1]{\ensuremath{\left(#1\right)}\xspace}

\newcommand{\doublebars}[1]{\ensuremath{\left\Vert#1\right\Vert}\xspace}

\newcommand{\complex}{\ensuremath{\mathbb{C}}\xspace}

\renewcommand{\j}{\ensuremath{\mathrm{j}}}

\newcommand{\inv}{\ensuremath{^{-1}}\xspace}

\newcommand{\trans}{\ensuremath{^{\mathrm{T}}}\xspace}
\newcommand{\ctrans}{\ensuremath{^{{*}}}\xspace}
\newcommand{\conj}{\ensuremath{^{\mathrm{c}}}\xspace} 

\newcommand{\pnorm}[2]{\ensuremath{\doublebars{#2}_{#1}}\xspace}

\newcommand{\normfro}[1]{\pnorm{\mathrm{F}}{#1}}

\newcommand{\distcgauss}[2]{\ensuremath{\mathcal{N}_{\complex}\parens{#1,#2}}\xspace} 

\newcommand{\arrayresponse}[1]{\ensuremath{\vec{a}\parens{#1}}\xspace}

\newcommand{\Nt}{\ensuremath{N_\mathrm{t}}\xspace} 
\newcommand{\Nr}{\ensuremath{N_\mathrm{r}}\xspace} 

\newcommand{\Lt}{\ensuremath{L_{\mathrm{t}}}\xspace} 

\newcommand{\symbvec}{\ensuremath{\vec{s}}\xspace}

\newcommand{\symbvecest}{\ensuremath{\hat{\vec{s}}}\xspace}

\newcommand{\Ns}{\ensuremath{N_\mathrm{s}}\xspace} 

\newcommand{\channel}{\ensuremath{\mat{H}}\xspace}

\newcommand{\pre}{\ensuremath{\mat{F}}\xspace}
\newcommand{\prebb}{\ensuremath{\pre_{\mathrm{BB}}}\xspace}
\newcommand{\prerf}{\ensuremath{\pre_{\mathrm{RF}}}\xspace}

\newcommand{\com}{\ensuremath{\mat{W}}\xspace}
\newcommand{\combb}{\ensuremath{\com_{\mathrm{BB}}}\xspace}
\newcommand{\comrf}{\ensuremath{\com_{\mathrm{RF}}}\xspace}

\newcommand{\noisevec}{\ensuremath{\vec{n}}\xspace}

\def\vn{{\vec{n}}}

\def\vs{{\vec{s}}}

\def\vw{{\vec{w}}}
\def\vx{{\vec{x}}}
\def\vy{{\vec{y}}}

\def\mH{{\mat{H}}}

\usepackage{xspace}
\usepackage[acronym,nogroupskip,nonumberlist,nopostdot]{glossaries}
\makeglossaries

\usepackage{xcolor}

\newacronym{snr}{SNR}{signal-to-noise ratio}
\newacronym{sinr}{SINR}{signal-to-interference-plus-noise ratio}
\newacronym{inr}{INR}{interference-to-noise ratio}
\newacronym{sir}{SIR}{signal-to-interference ratio}
\newacronym{sqr}{SQR}{signal-to-quantization-noise ratio}
\newacronym{sqnr}{SQNR}{signal-to-quantization-plus-noise ratio}
\newacronym{ian}{IAN}{interference as noise}
\newacronym{ber}{BER}{bit error rate}
\newacronym{pn}{PN}{pseudorandom noise}
\newacronym{bfsk}{BFSK}{binary frequency shift keying}
\newacronym{fh}{FH}{frequency-hopped}
\newacronym{fh-bfsk}{FH-BFSK}{frequency-hopped binary frequency shift keying}
\newacronym{crc}{CRC}{cyclic redundancy check}
\newacronym{isi}{ISI}{intersymbol interference}
\newacronym{dsss}{DSSS}{direct-sequence spread spectrum}
\newacronym{ofdm}{OFDM}{orthogonal frequency-division multiplexing}
\newacronym{ofdma}{OFDMA}{orthogonal frequency-division multiple access}
\newacronym{sdr}{SDR}{software-defined radio}
\newacronym{tx}{TX}{transmitter}
\newacronym{rx}{RX}{receiver}
\newacronym{fdd}{FDD}{frequency-division duplexing}
\newacronym{tdd}{TDD}{time-division duplexing}
\newacronym{fdma}{FDMA}{frequency-division multiple access}
\newacronym{tdma}{TDMA}{time-division multiple access}
\newacronym{sdma}{SDMA}{space-division multiple access}
\newacronym[plural=MPCs]{mpc}{MPC}{multipath component}
\newacronym{mui}{MUI}{multi-user interference}

\newacronym{qam}{QAM}{quadrature amplitude modulation}
\newacronym{mqam}{MQAM}{M-ary quadrature amplitude modulation}

\newacronym{ls}{LS}{least-squares}
\newacronym{lms}{LMS}{least mean squares}
\newacronym{rls}{RLS}{recursive least-squares}
\newacronym{rzf}{RZF}{regularized zero-forcing}
\newacronym{mmse}{MMSE}{minimum mean square error}
\newacronym{lmmse}{LMMSE}{linear minimum mean square error}
\newacronym{mse}{MSE}{mean square error}
\newacronym{fft}{FFT}{fast Fourier transform}
\newacronym{dft}{DFT}{discrete Fourier transform}
\newacronym{dtft}{DTFT}{discrete-time Fourier transform}
\newacronym{ctft}{CTFT}{continuous-time Fourier transform}
\newacronym{ml}{ML}{machine learning}
\newacronym[plural=NNs]{nn}{NN}{neural network}
\newacronym[plural=RNNs]{rnn}{RNN}{recurrent neural network}
\newacronym[plural=ADCs]{adc}{ADC}{analog-to-digital converter}
\newacronym[plural=DACs]{dac}{DAC}{digital-to-analog converter}
\newacronym[plural=FPGAs]{fpga}{FPGA}{field-programmable gate array}
\newacronym{evm}{EVM}{error vector magnitude}
\newacronym{enob}{ENOB}{effective number of bits}
\newacronym{zf}{ZF}{zero-forcing}
\newacronym{rv}{r.v.}{random variable}
\newacronym{omp}{OMP}{orthogonal matching pursuit}
\newacronym{svd}{SVD}{singular value decomposition}

\newacronym{sdp}{SDP}{semidefinite programming}
\newacronym{psd}{PSD}{positive semidefinite}
\newacronym{nsd}{NSD}{negative semidefinite}

\newacronym{agc}{AGC}{automatic gain control}
\newacronym{rf}{RF}{radio frequency}
\newacronym{los}{LOS}{line-of-sight}
\newacronym{nlos}{NLOS}{non-line-of-sight}
\newacronym{ple}{PLE}{path loss exponent}
\newacronym[plural=dB,firstplural=decibels (dB)]{db}{dB}{decibel}
\newacronym[plural=dBm,firstplural=decibel milliwatts (dBm)]{dbm}{dBm}{decibel milliwatts}
\newacronym{pa}{PA}{power amplifier}
\newacronym{lna}{LNA}{low noise amplifier}
\newacronym{cw}{CW}{continuous wave}
\newacronym{papr}{PAPR}{peak-to-average power ratio}
\newacronym{usrp}{USRP}{Universal Software Radio Peripheral}
\newacronym{irr}{IRR}{image rejection ratio}
\newacronym{lo}{LO}{local oscillator}
\newacronym{vm}{VM}{vector modulator}
\newacronym{mmwave}{mmWave}{millimeter wave}
\newacronym{thz}{THz}{terahertz}
\newacronym{eirp}{EIRP}{effective isotropic radiated power}

\newacronym{csma}{CSMA}{carrier-sense multiple access}
\newacronym{csmaca}{CSMA/CA}{carrier-sense multiple access with collision avoidance}
\newacronym{csmacd}{CSMA/CD}{carrier-sense multiple access with collision detection}
\newacronym{mac}{MAC}{medium access control}
\newacronym{phy}{PHY}{physical layer}
\newacronym{4g}{4G}{fourth generation}
\newacronym{lte}{LTE}{Long-Term Evolution}
\newacronym{4glte}{4G LTE}{\gls{4g} \gls{lte}}
\newacronym{5g}{5G}{fifth generation}
\newacronym{nr}{NR}{New Radio}
\newacronym{5gnr}{5G NR}{5G New Radio}
\newacronym{ieee}{IEEE}{Institute of Electrical and Electronics Engineers}
\newacronym{wifi}{Wi-Fi}{IEEE 802.11}
\newacronym{lan}{LAN}{local area network}
\newacronym{wlan}{WLAN}{wireless local area network}
\newacronym[plural=BSs]{bs}{BS}{base station}
\newacronym[plural=SBSs]{sbs}{SBS}{small-cell base station}
\newacronym[plural=FD-SBSs]{fdsbs}{FD-SBS}{\gls{fd}-enabled \gls{sbs}}
\newacronym[plural=MBSs]{mbs}{MBS}{macrocell base station}
\newacronym[plural=UEs]{ue}{UE}{user equipment}
\newacronym{ul}{UL}{uplink}
\newacronym{dl}{DL}{downlink}
\newacronym{qos}{QoS}{Quality of Service}
\newacronym{fcc}{FCC}{Federal Communications Commission}
\newacronym{iab}{IAB}{integrated access and backhaul}
\newacronym{fab}{FAB}{fixed access and backhaul}
\newacronym{hetnet}{HetNet}{heterogeneous network}

\newacronym{siso}{SISO}{single-input single-output}
\newacronym{mimo}{MIMO}{multiple-input multiple-output}
\newacronym{sumimo}{SU-MIMO}{single-user \gls{mimo}}
\newacronym{mumimo}{MU-MIMO}{multi-user \gls{mimo}}
\newacronym{bf}{BF}{beamforming}
\newacronym{ca}{CA}{constant amplitude}
\newacronym{ula}{ULA}{uniform linear array}
\newacronym{upa}{UPA}{uniform planar array}
\newacronym[\glslongpluralkey={angles of arrival}]{aoa}{AoA}{angle of arrival}
\newacronym[\glslongpluralkey={angles of departure}]{aod}{AoD}{angle of departure}
\newacronym{dof}{DoF}{degrees of freedom}
\newacronym{csi}{CSI}{channel state information}
\newacronym{csit}{CSIT}{\gls{csi} at the transmitter}
\newacronym{csir}{CSIR}{\gls{csi} at the receiver}
\newacronym{cs}{CS}{compressed sensing}

\newacronym{fd}{FD}{in-band full-duplex}
\newacronym{hd}{HD}{half-duplex}
\newacronym{si}{SI}{self-interference}
\newacronym{sic}{SIC}{self-interference cancellation}
\newacronym{soi}{SoI}{signal of interest}
\newacronym{asic}{A-SIC}{analog \acrlong{sic}}
\newacronym{dsic}{D-SIC}{digital \gls{sic}}
\newacronym{star}{STAR}{simultaneous transmit and receive}
\newacronym{warp}{WARP}{Wireless Open-Access Research Platform}
\newacronym{bfc}{BFC}{beamforming cancellation}
\newacronym{ipi}{IPI}{inter-panel-interference}
\newacronym{ipic}{IPIC}{inter-panel-interference cancellation}

\newacronym{qcqp}{QCQP}{quadratically-constrained quadratic programming}
\newacronym{cdf}{CDF}{cumulative density function}

\newacronym{elf}{ELF}{extremely low frequency}
\newacronym{slf}{SLF}{super low frequency}
\newacronym{ulf}{ULF}{ultra low frequency}
\newacronym{vlf}{VLF}{very low frequency}
\newacronym{lf}{LF}{low frequency}
\newacronym{mf}{MF}{medium frequency}
\newacronym{hf}{HF}{high frequency}
\newacronym{vhf}{VHF}{very high frequency}
\newacronym{uhf}{UHF}{ultra high frequency}
\newacronym{shf}{SHF}{super high frequency}
\newacronym{ehf}{EHF}{extremely high frequency}
\newacronym{thf}{THF}{tremendously high frequency}

\newacronym{wncg}{WNCG}{Wireless Networking and Communications Group}
\newacronym{linc}{LINC}{Laboratory of Informatics, Networks, and Communications}
\newacronym{ut}{UT Austin}{The University of Texas at Austin}
\newacronym{uiuc}{UIUC}{University of Illinois at Urbana-Champaign}
\newacronym{usc}{USC}{University of Southern California}
\newacronym{mit}{MIT}{Massachusetts Institute of Technology}
\newacronym{berkeley}{UC Berkeley}{University of California, Berkeley}
\newacronym{osu}{OSU}{Ohio State University}

\newacronym{mfm}{MFM}{\textsc{{MIMO} for Matlab}}
\newcommand{\mfm}{\gls{mfm}\xspace}
\newcommand{\matlab}{\textsc{Matlab}\xspace}

\newcommand{\mmwave}{\gls{mmwave}\xspace}
\newcommand{\thz}{\gls{thz}\xspace}
\newcommand{\mimo}{\gls{mimo}\xspace}
\newcommand{\siso}{\gls{siso}\xspace}
\newcommand{\mmse}{\gls{mmse}\xspace}

\newcommand{\ue}{\gls{ue}\xspace}

\newcommand{\rf}{\gls{rf}\xspace}

\newcommand{\ula}{\gls{ula}\xspace}
\newcommand{\upa}{\gls{upa}\xspace}

\newcommand{\figref}[1]{\figurename~\ref{#1}}

\usepackage{verbatim}
\newcommand{\varray}{\texttt{array}\xspace}
\newcommand{\vchannel}{\texttt{channel}\xspace}
\newcommand{\vdevice}{\texttt{device}\xspace}
\newcommand{\vlink}{\texttt{link}\xspace}
\newcommand{\vnetwork}{\texttt{network\_mfm}\xspace}
\newcommand{\vpathloss}{\texttt{path\_loss}\xspace}
\newcommand{\vreceiver}{\texttt{receiver}\xspace}
\newcommand{\vreceiverh}{\texttt{receiver\_hybrid}\xspace}
\newcommand{\vtransmitter}{\texttt{transmitter}\xspace}
\newcommand{\vtransmitterh}{\texttt{transmitter\_hybrid}\xspace}

\begin{document}
	
%
\title{\textsc{MIMO for Matlab}: A Toolbox for\\Simulating MIMO Communication Systems}

\author{
    \IEEEauthorblockN{Ian~P.~Roberts}%
    \IEEEauthorblockA{\texttt{\url{https://mimoformatlab.com}}}
}

\maketitle




\begin{abstract}
We present \mfm, a toolbox for \matlab that aims to simplify the simulation of \mimo communication systems research while facilitating reproducibility, consistency, and community-driven customization.
\mfm offers users an object-oriented solution for simulating a variety of \mimo systems including sub-6 GHz, massive \mimo, millimeter wave, and terahertz communication.
Out-of-the-box, \mfm supplies users with widely used channel and path loss models from academic literature and wireless standards; if a particular channel or path loss model is not provided by \mfm, users can create custom models by following a few simple rules.
The complexity and overhead associated with simulating networks of multiple devices can be significantly reduced with \mfm versus raw \matlab code, especially when users want to investigate various channel models, path loss models, precoding/combining schemes, or other system-level parameters.
\mfm's heavy-lifting to automatically collect and distribute channel state information, aggregate interference, and report performance metrics relieves users of otherwise tedious tasks and instills confidence and consistency in the results of simulation.
The use-cases of \mfm vary widely from networks of hundreds of devices; to simple point-to-point communication; to serving as a channel generator; to radar, sonar, and underwater acoustic communication.
\end{abstract}





\glsresetall


\section{Overview} \label{sec:overview}

Research and education on \mimo communication systems are built on linear equations of the form
\begin{align}
\symbvecest = \sqrt{P} \cdot G \cdot \com\ctrans \channel \pre \symbvec + \com\ctrans \noisevec \label{eq:batman}
\end{align}
sometimes termed \textit{symbol-level} or \textit{single-letter} formulations \cite{heath_lozano}.
To communicate a symbol vector $\symbvec$ over some channel matrix $\channel$, a transmitter applies power $P$ and a precoding matrix $\pre$ while a receiver applies a combining matrix $\com$ to recover an estimate $\symbvecest$ of the symbol vector $\symbvec$.
Along the way, path loss $1/G^2$ weakens the transmitted signal and additive noise $\noisevec$ corrupts the received signal.
While these linear models greatly simplify the sophistication of today's communication systems, simulating \mimo concepts and research can become prohibitively complex and overwhelming when a network grows to even a moderate size.
The variety of channel models and precoding/combining strategies used in \mimo communication, along with enforcing common normalizations, can further complicate simulation and introduce the potential for mistakes and inconsistency.

This has motivated us to create \mfm, a toolbox for simulating \mimo communication systems \cite{mfm}. 
\mfm is written in an object-oriented fashion and comes with a collection physical layer tools including a variety of channel models, path loss models, transmitters, receivers, and antenna arrays that can be used for sub-6 GHz, \mmwave, \thz, and beyond.
It supports both fully-digital and hybrid digital/analog transceivers with options for limited phase and amplitude control as well as fully- and partially-connected analog beamforming networks.
In addition to its applications in research, \mfm can act as an educational tool to help students understand, experiment, and visualize \mimo communication.

\mfm is a physical layer toolbox that has support from the antenna/spatial domain all the way up to a network of users.
By design, \mfm has been created to be used at any level within its capabilities.
For instance, \mfm can be used at its lowest level for antenna array research or to draw channel realizations from a particular model.
At its highest level, \mfm can be used to simulate a network of many users, automatically aggregating interference inflicted at each device by one another.
In between, simple point-to-point communication can be simulated, allowing users to develop, implement, and evaluate novel precoding and combining schemes.

\begin{figure}
    \centering
    \includegraphics[width=\linewidth,height=0.5\textheight,keepaspectratio]{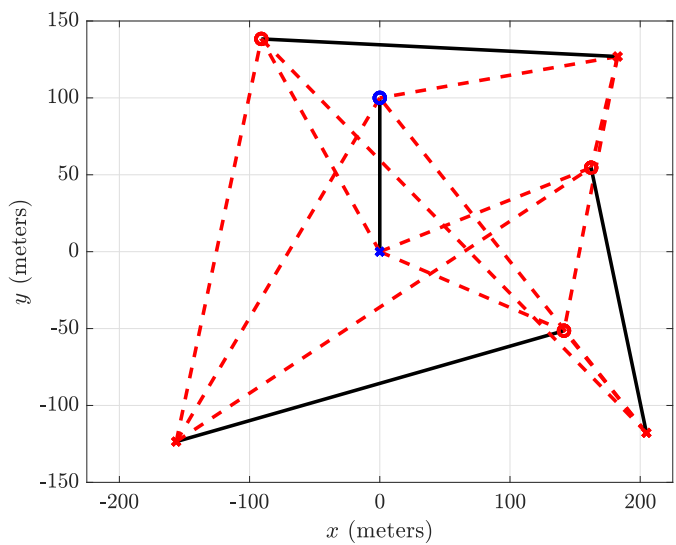}
    \caption{A network of eight devices scattered in space simulated in \mfm. Four transmit-receive pairs (shown as $\times$'s and $\circ$'s, respectively) use the same time-frequency resource in their attempt to individually communicate.}
    \label{fig:network}
\end{figure}


\mfm is freely available for use under the MIT license and can be setup in \matlab in minutes.
It has been completely documented in-line, which can be accessed using \matlab's \texttt{help} function.
In addition, the \mfm website has extensive amounts of documentation, examples, and tutorials, which will continue to be updated.
Along with the official \mfm contents, users are encouraged to develop and share their own channel models, path loss models, etc.~to extend \mfm's capabilities and 
We hope \mfm will act as a platform that facilitates consistency and reproducibility of \mimo research and accelerates its development.
We are interested in tracking the reach it has and applications it serves to better improve \mfm in the future.
If you use \mfm, please cite this paper and also the package itself. 

\begin{figure}
    \centering
    \includegraphics[width=\linewidth,height=0.3\textheight,keepaspectratio]{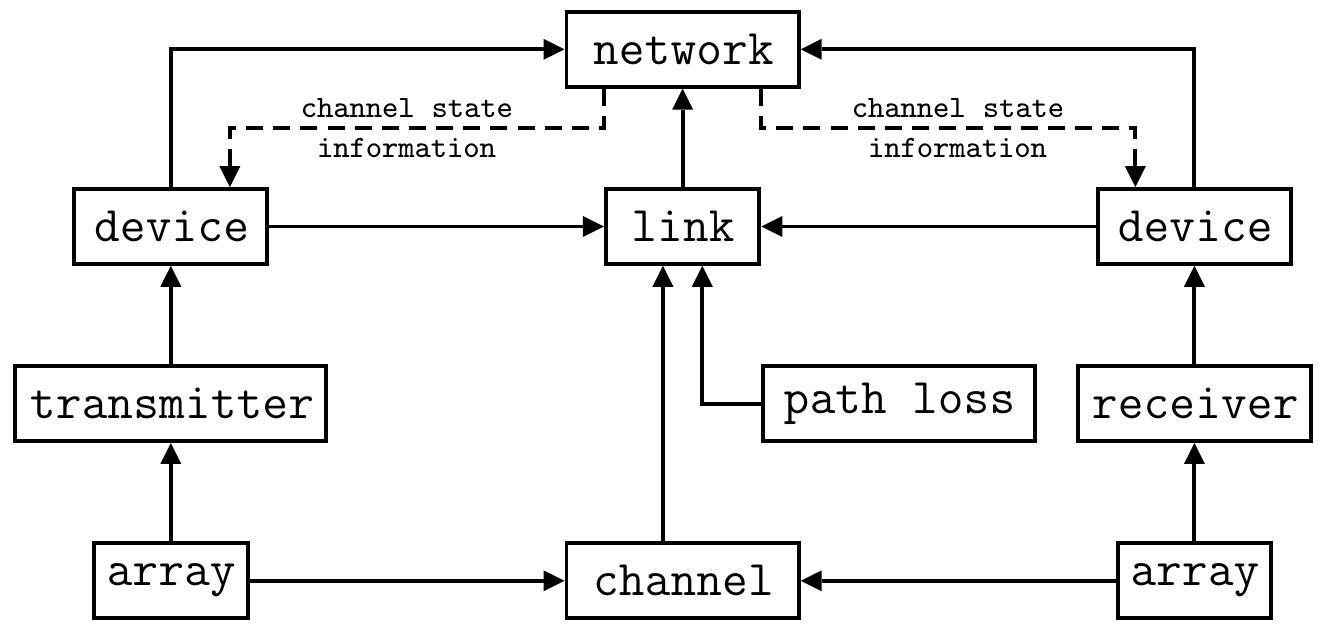}
    \caption{Example architecture of an \mfm script simulating point-to-point communication between a transmitting device and receiving device.}
    \label{fig:oop}
\end{figure}

\mfm executes physical layer signal processing at the symbol level, abstracting out the waveforms that carry those symbols as is commonly done in \mimo literature.
The object-oriented structure of \mfm can be summarized as in \figref{fig:oop}, though select object(s) can be useful on their own.
At the lowest level of \mfm are antenna arrays, channel models, and path loss models.
Transmitters and receivers leverage antenna arrays to execute precoding and combining, respectively.
Devices, which can have transmit and/or capability, are connected to one another via links, which capture propagation via channel and path loss models.
The collection of devices and the links connecting them comprise a network.
A network instance in \mfm captures devices operating on the same time-frequency resource, meaning all transmitting devices in a network can inflict interference onto all receiving devices.
As such, it is up to users of \mfm to properly choose devices present in the network to capture time- or frequency-division.

\mfm is a collection of MATLAB scripts that can be used together, to varying degrees, to simulate \mimo communication systems.
The \mfm framework simplifies generating channels/network realizations, executing precoding and combining strategies, and evaluating communication system performance.
With \mfm, users can focus their attention on the aspects of \mimo communication that are \textit{relevant to them} since \mfm can handle the rest.
For example, users interested in creating \mimo precoding and combining strategies may want to examine their strategies across many channel and path loss models.
\mfm can enable such by providing a collection of common channel and path loss models, which can be used interchangeably network-wide with ease.
In addition, \mfm's heavy-lifting can relieve users of the headache associated with tasks such as computing interference and collecting channel state information, which grow daunting and overwhelming with networks of moderate size.

As mentioned, \mfm can be used to varying degrees.
For beamforming and array-related work, users may only need \mfm's antenna array object.
Those interested in using \mfm to generate channel realizations can use the antenna array and channel objects.
To use \mfm to simulate point-to-point \mimo communication---perhaps to experiment with precoding/combining schemes---can use \mfm at the link level.
To capture the impacts of interference, users can use it at the network level.
We imagine \mfm could be used for a variety of applications beyond strictly \mimo research including stochastic geometry, joint communication and radar, underwater acoustic communication, sonar, machine learning in communications, and satellite communication.
While \mfm is currently strictly a physical layer toolbox, other areas of research, such as on scheduling, could leverage \mfm to avoid the headache associated with implementing physical layer communication network-wide. 

By using \mfm as a common framework across the research community, researchers can share their \mfm scripts and objects to facilitate reproducibility, broadening the impact of their work, and instilling confidence in their results.
Thanks to its object-oriented design, \mfm objects created by users can be easily shared and implemented across the research community.
\mfm was designed to accommodate customizations and expansions that a user sees fit.
For example, if a particular channel model that a user needs is not provided in \mfm, users can create their own by following a few simple rules.
Once created, the custom channel model can be easily shared and then incorporated into \mfm by others across the research community.
If particular additions to \mfm are widely used, there are avenues for them to be incorporated into future versions of \mfm.

\section{Low Level Objects and Usage} \label{sec:low-level}

At \mfm's lowest level are its antenna arrays, channel models, and path loss models.

\subsection{Antenna Arrays}
From which the term \mimo takes its name, let us begin by outlining support for antenna arrays \cite{balanis}.
\mfm's \varray object is used to represent antenna arrays, which can be constructed as \glspl{ula}, \glspl{upa}, or other arbitrary array the user wishes.
Arrays can be constructed in a few ways:
\begin{itemize}
    \item \verb|a = array.create()| creates an empty array with no elements, after which elements can be added to the array.
    \item \verb|a = array.create(N)| creates a half-wavelength \ula with \texttt{N} elements.
    \item \verb|a = array.create(M,N)| creates a half-wavelength \upa with \texttt{M} rows of \texttt{N} elements.
\end{itemize}
Arrays can be rotated and translated as desired and individual array elements can be added or removed.
To simulate a typical \mimo communication system, creating an \varray using the \ula or \upa method will often suffice without significant modification, especially when the channel model used is independent of the array geometry (e.g., a Rayleigh-faded channel).
Currently, \mfm assumes isotropic elements, though support for more practical element patterns will likely be included in future versions.


Some array configurations---such as half-wavelength uniform linear and planar arrays---have well-known expressions for their response as a function of direction.
While such array configurations also happen to be the most commonly used, \mfm supports arbitrary antenna arrays.
In other words, \mfm does not restrict the type of arrays a user can construct.
In general, array elements can be placed arbitrarily in 3-D in units of carrier wavelengths $\lambda$, which makes the array behavior agnostic of the carrier frequency at which it operates.
To handle arbitrary array construction, \mfm computes the array response based on the relative positioning of the array elements.

The relative phase shift experienced by the $i$-th array element located at some relative $(x_i,y_i,z_i)$ due to a plane wave in the direction $(\theta,\phi)$ is
\begin{align}
a_i\parens{\theta,\phi} = \exp\parens{\j \cdot \frac{2 \pi}{\lambda} \cdot \zeta\parens{x_i,y_i,z_i,\theta,\phi}} \label{eq:spiderman}
\end{align}
where $\lambda$ is the carrier wavelength and
\begin{align}
\zeta\parens{x,y,z,\theta,\phi} = {x \sin \theta \cos \phi + y \cos \theta \cos \phi + z \sin \phi}
\end{align}
The array response vector is constructed by collecting the relative phase shift seen by each of the array's $N$ elements as
\begin{align}
\arrayresponse{\theta,\phi} = 
[
a_1\parens{\theta,\phi}, 
a_2\parens{\theta,\phi}, 
\dots, 
a_{N}\parens{\theta,\phi}
]\trans
\end{align}
To obtain the array response of an array \verb|a| in a particular azimuth \verb|theta| and elevation \verb|phi| in \mfm, one simply needs to call
\begin{verbatim}
v = a.get_array_response(theta,phi)
\end{verbatim}
\mfm will automatically populate the array response vector \verb|v| based on the array geometry.

To weight the $N$ elements of an antenna array \verb|a|, one can use \verb|a.set_weights(w)|, where \texttt{w} is a vector of $N$ complex weights.
Note that the weights contained in \texttt{w} are applied as is and are not conjugated beforehand.
This can be described mathematically by stating that the gain of an array with weights $\vw$ in the direction $\parens{\theta,\phi}$ is
\begin{align}
g\parens{\theta,\phi} = \vw\trans \arrayresponse{\theta,\phi}
\end{align}
where $(\cdot)\trans$ denotes transpose (not conjugate transpose).
Therefore, to so-called conjugate beamform (i.e., matched filter) in the direction of $\parens{\theta,\phi}$, one would take $\vw = \arrayresponse{\theta,\phi}\conj$, where $(\cdot)\conj$ denotes element-wise conjugation.
To achieve this in \mfm, this would simply be
\begin{verbatim}
v = a.get_array_response(theta,phi)
w = conj(v)
a.set_weights(w)
\end{verbatim}

\begin{figure}
    \centering
    \includegraphics[width=\linewidth,height=0.2\textheight,keepaspectratio]{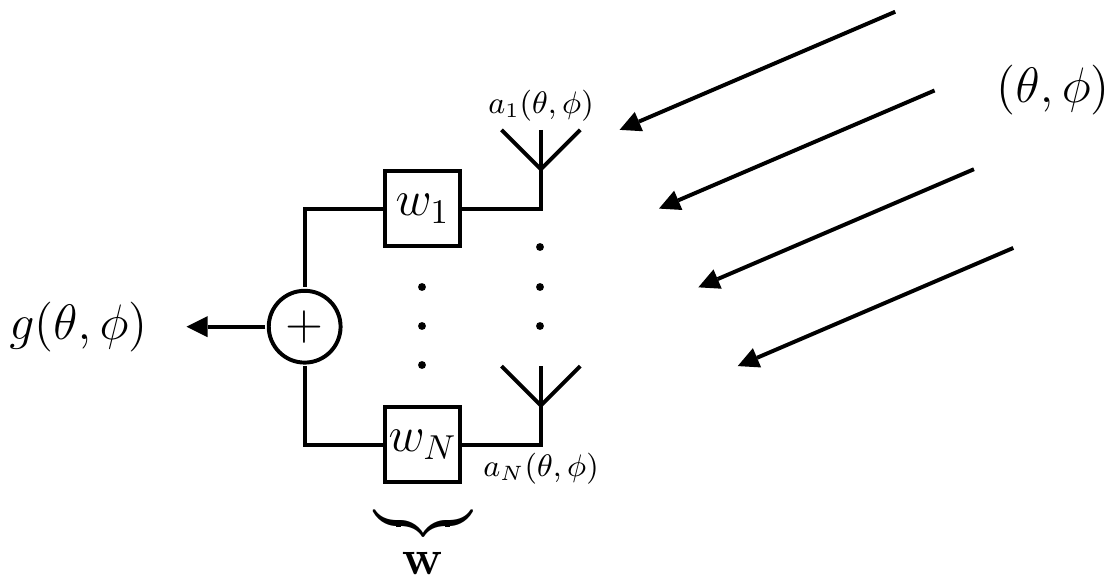}
    \caption{The gain of a weighted array in a direction $(\theta,\phi)$.}
    \label{fig:array-weighted-response}
\end{figure}

To evaluate the complex gain \verb|g| achieved by a weighted array \verb|a| in the direction (\texttt{theta},\texttt{phi}), one could use the following.
\begin{verbatim}
g = a.get_array_gain(theta,phi)
\end{verbatim}

As will be discussed, the precoding and combining executed by \mfm does not use this beamforming feature of \varray objects, even hybrid digital/analog precoding and combining architectures.
Instead, \mfm executes precoding and combining at the \vtransmitter and \vreceiver objects, respectively.


\subsection{Channel Models}
The \vchannel objects in \mfm are used to capture the over-the-air mixing that takes place across transmit antennas and receive antennas, leading to a channel matrix $\mH$.
Following convention in \mimo literature, channel matrices $\mH$ in \mfm are always of size $\Nr \times \Nt$, where $\Nr$ antennas are at the receiver and $\Nt$ antennas are at the transmitter.
Currently, \mfm only supports these frequency-flat channels but work is ongoing to extend its support to frequency-selective ones.
It is important to keep in mind that while \mfm has \mimo in its name, it also supports \siso scenarios to an extent.
\mfm supports a variety of channel models including the Rayleigh-faded channel, \gls{los} channel, ray/cluster channel (extended Saleh-Valenzuela model \cite{heath_overview_2016}), and spherical-wave channel \cite{spherical_2005}.


To create a Rayleigh-faded channel object \verb|c|, for instance, one can simply use
\begin{verbatim}
c = channel.create('Rayleigh')
\end{verbatim}
Other channels can be created similarly.
Setup of a channel \verb|c| begins by declaring the propagation velocity (e.g., $3 \times 10^8$) and carrier frequency using
\begin{verbatim}
c.set_propagation_velocity(vel)
c.set_carrier_frequency(fc)
\end{verbatim}
While \mfm was created for conventional electromagnetic-based wireless communication, affording users the ability to set the propagation velocity may lend \mfm support to other fields such as underwater acoustic communication where the propagation velocity of sound in the ocean is often taken to be around $1.5 \times 10^3$ m/s, for example.
Note that setting the carrier frequency will automatically update the channel's carrier wavelength according to its propagation velocity.

The next necessary step is informing the channel of the transmit and receive arrays between which it lives, which can be accomplished by
\begin{verbatim}
c.set_arrays(atx,arx)
\end{verbatim}
where \verb|atx| and \verb|arx| are the transmit and receive array objects, respectively.
Informing \vchannel objects of the transmit and receive arrays dictates the size of the to-be-realized channel matrix $\mH$ and also provides geometric channel models---such as the \gls{los} channel and ray/cluster channel---with access to the array responses.
Each channel model will have unique setup steps before it can becomes useful.


Once a \vchannel object \texttt{c} has been created and properly set up, a realization of the channel is merely one line of code.
\begin{verbatim}
H = c.realization()
\end{verbatim}
This is especially convenient for Monte Carlo simulations, where channel realizations are placed within a loop, as below.
\begin{verbatim}
for i = 1:N
    ...
    H = c.realization()
    ...
end
\end{verbatim}
Any stochastics associated with the channel model will be redrawn from their respective distributions when constructing the channel matrix on each realization.
Users can create custom channel models that are compatible with \mfm by following a few simple rules; more information can be found on the \mfm website.

\subsection{Path Loss Models}
While channel models capture the small-scale mixing between transmit and receive antennas, path loss models capture the large-scale gain $G$ between a transmitter and receiver (e.g., due to propagation loss, shadowing, blockage, etc.).
Path loss models can be used on their own to realize values of $G$ directly or in the point-to-point and network settings.
A number of path loss models exist in \mfm including free-space path loss (with and without log-normal shadowing) and two-slope path loss.
More path loss models are continuing to be added to \mfm.
Creating a free-space path loss object \verb|p| can be achieved via
\begin{verbatim}
p = path_loss.create('FSPL')
\end{verbatim}
After setting its propagation velocity and carrier frequency, the path loss exponent \verb|ple| can be set as
\begin{verbatim}
p.set_path_loss_exponent(ple)
\end{verbatim}
The distance of the path \verb|d| can be set using
\begin{verbatim}
p.set_distance(d)
\end{verbatim}
From there, the path loss can be realized using
\begin{verbatim}
L = p.realization()
\end{verbatim}
which will return a power loss \verb|L| according to the free-space formula
\begin{align}
L\inv = G^{2} = \left( \frac{\lambda}{4\pi} \right)^{2} \times \left( \frac{1}{d} \right)^{\eta}
\end{align}
where $\eta$ is the path loss exponent and $d$ is the distance of the path.
More complicated path loss models may require additional setup, and those involving stochastics (e.g., with shadowing) may realize a random path loss on each realization.

\section{Link-Level Usage} \label{sec:link-level}

\subsection{Transmitters}
A transmitter in \mfm is captured by the \vtransmitter object and its subclasses.
A \vtransmitter can be created via
\begin{verbatim}
tx = transmitter.create()
\end{verbatim}
By default, a \vtransmitter object employs fully-digital precoding, but \mfm also supports hybrid digital/analog precoding. 
The symbol vector departing a fully-digital transmitter follows the form
\begin{align}
\vx = \sqrt{P} \cdot \pre \symbvec \label{eq:hulk}
\end{align}
where $P$ reflects the transmit power applied to a symbol vector $\vs$ having undergone precoding by a matrix $\pre$.
The main properties of a \vtransmitter include its antenna array, transmit power, precoding matrix (i.e., $\pre$), precoding power budget, transmit symbol (i.e., $\vs$), channel state information, and symbol bandwidth (i.e., $B$).
A \vtransmitter's properties can be set using various \verb|set| commands.
For example, to set the transmit power of a transmitter \verb|tx|, one can simply use
\begin{verbatim}
tx.set_transmit_power(P,'dBm')
\end{verbatim}
where \verb|P| is the transmit power in dBm.

To limit the power associated with precoding, \mfm supports a precoding power budget, which takes on the form
\begin{align}
\normfro{\pre}^2 \leq E
\end{align}
where $E$ is the precoding power budget.
By default, \mfm sets the precoding power budget to $E = \Ns$.

The precoding matrix can be set using
\begin{verbatim}
tx.set_precoder(F)
\end{verbatim}
where \verb|F| is an $\Nt \times \Ns$ precoding matrix, or using other methods as we will discuss shortly.

\mfm supports hybrid digital/analog precoding via its \vtransmitterh object, which is a subclass of the \vtransmitter object, meaning it inherits all of the properties and functions discussed so far.
The symbol vector departing a hybrid transmitter follows the form
\begin{align}
\vx = \sqrt{P} \cdot \prerf \prebb \symbvec \label{eq:iron-man}
\end{align}
where a digital precoding matrix $\prebb$ followed by an analog precoding matrix $\prerf$ are applied to symbol vector $\vs$.
A hybrid digital/analog transmitter can be created by including the \verb|'hybrid'| specifier when creating a transmitter.
\begin{verbatim}
tx = transmitter.create('hybrid')
\end{verbatim}
The number of \rf chains, phase and amplitude resolution of analog beamforming, and connected-ness of the analog beamforming network can all be configured as desired.

Once setup, the hybrid transmitter's digital and analog precoders can be set via 
\begin{verbatim}
tx.set_precoder_digital(F_BB)
tx.set_precoder_analog(F_RF)
\end{verbatim}
where \verb|F_BB| is an $\Lt \times \Ns$ digital precoding matrix and \verb|F_RF| is an $\Nt \times \Lt$ analog precoding matrix.

Methods to explicitly set the precoding matrices have been discussed.
However, this is not always a very attractive approach, especially since it can severely undermine the advantages of \mfm's object-oriented design.
This motivates setting a transmitter's precoder(s) via
\begin{verbatim}
tx.configure_transmitter(strategy)
\end{verbatim}
where \verb|strategy| is a string specifying the strategy/method to use when designing the transmitter's precoder(s).
For example, for eigen-based precoding, one can use
\begin{verbatim}
tx.configure_transmitter('eigen')
\end{verbatim}
which will automatically use the transmitter's channel state information to design its precoder.
This string-based way to specify a transmit strategy is particularly useful since it keeps main simulation scripts free of the linear algebra involved in precoder design, allows users to easily switch between strategies, and makes setting precoders network-wide much more manageable.
Users can add custom transmit strategies with a few simple steps.

\subsection{Receivers}
A receiver in \mfm is captured by the \vreceiver object and its subclasses.
A \vreceiver can be created via
\begin{verbatim}
rx = receiver.create()
\end{verbatim}
Like the transmitter, a \vreceiver object employs fully-digital precoding by default, though hybrid digital/analog receivers are supported.
The estimated symbol vector output by a fully-digital receiver follows the form
\begin{align}
\symbvecest = \com\ctrans (\vy + \vn) \label{eq:captain-america}
\end{align}
where a combining matrix $\com$ is applied to the signal vector $\vy$ impinging the receive array plus noise $\vn$.
The main properties of a \vreceiver include its antenna array, combining matrix (i.e., $\com$), receive symbol (i.e., $\symbvecest$), noise power spectral density (i.e., $N_0$ or $\sigma_{\mathrm{n}}^2$), channel state information, and symbol bandwidth (i.e., $B$).
Like the \vtransmitter, a \vreceiver object's properties can be set using its various \verb|set| commands.

\mfm models noise as being additive, i.i.d. Gaussian across receive antennas.
The noise vector $\vn$ is drawn from the complex Gaussian distribution as
\begin{align}
\vn \sim \distcgauss{\mat{0}}{\sigma_{\mathrm{n}}^2 \cdot \mat{I}}
\end{align}
where $\sigma_{\mathrm{n}}^2$ is the average noise energy per symbol (joules) (i.e., the noise power spectral density).
To set the noise energy per symbol, we can use
\begin{verbatim}
rx.set_noise_power_per_Hz(psd,'dBm_Hz')
\end{verbatim}
where \verb|psd| is the noise power spectral density is in dBm/Hz. 



A receiver's combiner can be set via
\begin{verbatim}
rx.set_combiner(W)
\end{verbatim}
where \verb|W| is an $\Nr \times \Ns$ combining matrix.
Like the transmitter, receiver combining strategies can be specified using strings like \verb|'eigen'| or \verb|'mmse'| with \verb|rx.configure_receiver(strategy)| rather than setting the combining matrix explicitly.


Like with transmission, \mfm supports hybrid digital/analog receivers via its \vreceiverh object, which is a subclass of the \vreceiver object, meaning it inherits all of the \vreceiver properties and functions discussed so far.
The receive symbol output by a hybrid receiver takes the form
\begin{align}
\symbvecest = \combb\ctrans \comrf\ctrans (\vy + \vn)
\end{align}
where an analog combining matrix followed by a digital combining matrix are applied to a received signal vector $\vy$ plus noise $\vn$.
A hybrid receiver can be created by including the \verb|'hybrid'| specifier when creating a receiver.
\begin{verbatim}
rx = receiver.create('hybrid')
\end{verbatim}
The settings pertinent to a hybrid digital/analog receiver can be set with the same functions of a hybrid digital/analog transmitter.
The number of \rf chains, connected-ness of the hybrid receiver, and constraints of analog combining can all be configured in the same fashion as with the hybrid transmitter.

\subsection{Devices}
The \vdevice object, as its name suggests, represents a wireless communications terminal such as a \ue, base station, and the like.
A \vdevice object having only transmit \textit{or} receive capability will contain a \vtransmitter \textit{or} \vreceiver, respectively.
A \vdevice object that has both transmit \textit{and} receive capability---i.e., a transceiver---will be comprised of both a \vtransmitter and \vreceiver.

An \vdevice can be created via
\begin{verbatim}
d = device.create(type)
\end{verbatim}
where \verb|type| is either \verb|'transmitter'|, \verb|'receiver'|, or \verb|'transceiver'| (default).
%
It can be placed in 3-D space by setting its coordinate via 
\begin{verbatim}
d.set_coordinate(x,y,z)
\end{verbatim}
where \verb|x|, \verb|y|, and \verb|z| are Cartesian coordinates in meters.
The location of the \vdevice will be essential for geometry-dependent path loss and channel models in addition to visualization.
Other device parameters can be set straightforwardly using a variety of \verb|set| commands.

In many ways, the \vdevice object acts as a proxy for configuring and interfacing with its transmitter and/or receiver.
As such, the \vdevice is supplied with a number of \textit{passthrough} functions that make directly interfacing with its transmitter and/or receiver simpler.
For instance, the passthrough function 
\begin{verbatim}
d.set_transmit_power(P,'dBm')
\end{verbatim}
will set the transmit power of the device's transmitter, rather than the user needing to do 
\begin{verbatim}
d.transmitter.set_transmit_power(P,'dBm')
\end{verbatim}

Declaring which other device \verb|dev| a given device \verb|d| should transmit to (i.e., the \textit{destination} device) is accomplished via
\begin{verbatim}
d.set_destination(dev)
\end{verbatim}
where \verb|dev| is a device with receive capability.
Likewise, declaring the device it should receive from (i.e., \textit{source} device) is accomplished via
\begin{verbatim}
d.set_source(dev)
\end{verbatim}
where \verb|dev| is a device with transmit capability.
This source-destination concept is pertinent to particular use-cases of \mfm, particularly at its link and network levels, which will be discussed shortly, though also can be used in scenarios outside of such.


\subsection{Links}

\begin{figure}
    \centering
    \includegraphics[width=\linewidth,height=\textheight,keepaspectratio]{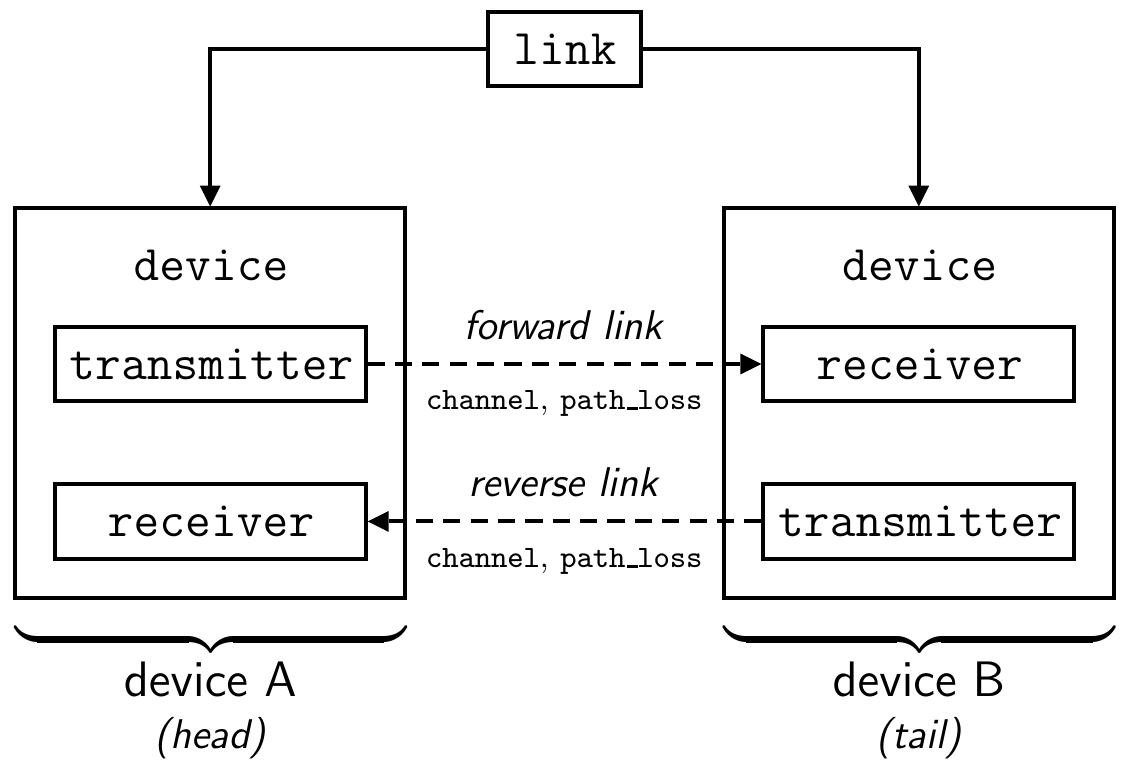}
    \caption{A link established between two transceivers.}
    \label{fig:link}
\end{figure}


Between any two devices sharing the same radio resources exists a channel matrix and path loss connecting them.
\mfm employs exactly that in its \vlink object used to connect a pair of \vdevice objects, with the caveat that one of the devices must have transmit capability and the other have receive capability; otherwise the physical connection (or \textit{link}) between the two devices would be immaterial.
Examining our familiar \mimo formulation, we can see that \mfm uses a \vlink to capture the channel matrix $\mH$ and large-scale gain $G$ due to path loss.

Suppose there exist two devices \verb|d1| and \verb|d2|, where \verb|d1| has transmit capability and \verb|d2| has receive capability.
Note that one or both devices could be transceivers.
A \vlink connecting these two devices is created via
\begin{verbatim}
lnk = link.create(d1,d2);
\end{verbatim}
By convention, the first device (\verb|d1| in this case) is called the \textit{head} while the second device (\verb|d2|) is called the \textit{tail}.
The head always has transmit capability and the tail always has receive capability.

Since both \verb|d1| and \verb|d2| could be transceivers, a link may exist between the two \vdevice objects in both directions, as illustrated by \figref{fig:link}. 
We refer to the link from the head to the tail as the \textit{forward link} and from the tail to the head as the \textit{reverse link}.
To handle cases when the forward and reverse links are \textit{symmetric} (or reciprocal), a single \vlink object contains both the forward and reverse links.
A \vlink object, therefore, has two \vchannel objects (a forward channel and reverse channel) and two \vpathloss objects (forward path loss and reverse path loss).

The channel and path loss models used on the forward and reverse links of a \vlink object \texttt{lnk} can be set using
\begin{verbatim}
lnk.set_channel(chan_fwd,chan_rev)
lnk.set_path_loss(path_fwd,path_rev)
\end{verbatim}
where \verb|chan_fwd| and \verb|chan_rev| are \verb|channel| objects and \verb|path_fwd| and \verb|path_rev| are \verb|path_loss| objects.

\section{Network-Level Usage} \label{sec:network-level}

At the highest level of \mfm's object-oriented structure is the \vnetwork object, which houses \vdevice objects and the \vlink objects connecting them.
Recall that the \vlink object represents a \textit{physical} connection between two devices rather than a \textit{communication} link.
A \vnetwork object can be created simply via
\begin{verbatim}
net = network_mfm.create()
\end{verbatim}
Currently, a network in \mfm captures scenarios where all devices present in the network share the same time-frequency resource, meaning some degree of interference will be inflicted onto each receiver in the network by each transmitter.


Suppose we have two \vdevice objects \verb|dtx| and \verb|drx|, where \verb|dtx| is a transmitting device and \verb|drx| is a receiving device, both of which have already been set up as necessary.
To inform the network that \verb|dtx| should transmit to \verb|drx| and that \verb|drx| should receive from \verb|dtx|, the following command is used
\begin{verbatim}
net.add_source_destination(dtx,drx);
\end{verbatim}
which adds \verb|dtx| and \verb|drx| as a \textit{source-destination pair}, \verb|dtx| being the source and \verb|drx| being the destination.
There are multiple ways to add devices to a network; this is merely the most useful.

To establish a physical connection (i.e., channel and path loss) between \textit{each} transmitting device and \textit{each} receiving device in the network, links can be added manually, though this is very cumbersome even for a small networks.
Fortunately, \mfm comes with a more convenient way of automatically populating links between pairs of devices via
\begin{verbatim}
net.populate_links_from_source_destination()
\end{verbatim}
which will populate all links from \textit{each} source device to \textit{each} destination device.
Recall that since all devices in an \mfm network share the same radio resources, each transmitting device will impose interference on those it does not intend to transmit to, meaning the number of links is equal to the number of source-destination pairs squared.

To specify the channel and path loss models used on all links in the network, we can invoke
\begin{verbatim}
net.set_channel(c)
net.set_path_loss(p)
\end{verbatim}
where \verb|c| and \verb|p| are \vchannel and \vpathloss objects, respectively.
\mfm offers the user the convenience of setting various system parameters network-wide instead of setting them at each device one-by-one, such as the carrier frequency, noise power, etc.


Once it has been properly setup, invoking a realization of an entire network \verb|net| is achieved with a single line
\begin{verbatim}
net.realization()
\end{verbatim}
which realizes all channels and path loss models in the network.
To collect and distribute channel state information across a network \verb|net|, use
\begin{verbatim}
net.compute_channel_state_information()
net.supply_channel_state_information()
\end{verbatim}
which can be used by devices to automatically configure themselves based on their transmit and receive strategies.
To configure \textit{all} transmitters and receivers across the network using string-specified transmit and receive strategies, use, for example, 
\begin{verbatim}
net.configure_transmitter('eigen')
net.configure_receiver('mmse')
\end{verbatim}
which transmits using eigen-beamforming and receives in an \mmse fashion.

With a network realized and its devices configured, the received signals can be computed via
\begin{verbatim}
net.compute_received_signals()
\end{verbatim}
which will \textit{automatically} aggregate interference caused by other transmitting devices in the network.

To report the mutual information (under Gaussian signaling) achieved between a particular pair of devices \verb|dev_1| and \verb|dev_2|, use
\begin{verbatim}
mi = net.report_mutual...
_information(dev_1,dev_2)
\end{verbatim}
which will automatically account for interference caused by other transmitting devices in the network.
To report the symbol estimation error achieved between a particular pair of devices \verb|dev_1| and \verb|dev_2|, use
\begin{verbatim}
[err,nerr] = net.report_symbol...
_estimation_error(dev_1,dev_2)
\end{verbatim}
where the first return value \verb|err| is the absolute symbol estimation error
$\lVert \hat{\mathbf{s}} - \mathbf{s} \rVert_{2}^{2}$
and the second return value \verb|nerr| is the symbol estimation error normalized to the transmit symbol energy defined as
\begin{align}
\frac{\lVert \hat{\mathbf{s}} - \mathbf{s} \rVert_{2}^{2}}{\lVert \mathbf{s} \rVert_{2}^{2}}
\end{align}

\begin{figure}
    \centering
    \includegraphics[width=\linewidth,height=\textheight,keepaspectratio]{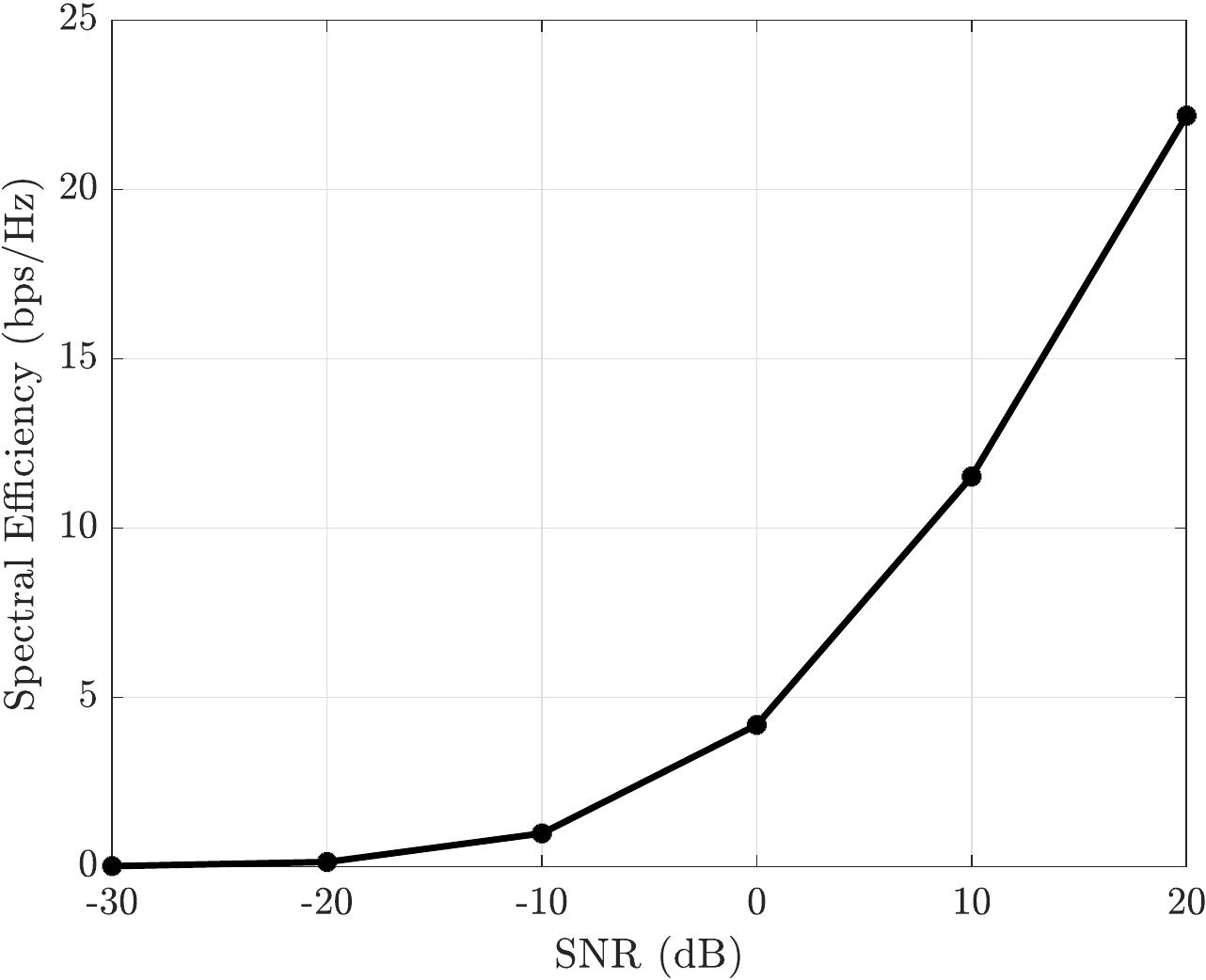}
    \caption{The spectral efficiency of a point-to-point Rayleigh-faded network as a function of \gls{snr} simulated using \mfm.}
    \label{fig:spec-eff}
\end{figure}

\section{Conclusion} \label{sec:conclusion}

\mfm aims to improve the reproducibility of \mimo research by providing a common framework for researchers to use when implementing their work. 
Out-of-the-box, \mfm is equipped with a variety of widely used channel and path loss models. 
Custom objects that integrate with \mfm can be created independently by users by following a few simple rules. 
These third-party customizations to \mfm can then be shared within the research community for others to use and potentially integrated into \mfm itself.
Beyond research, \mfm's potential also lay in educating students on \mimo and general wireless, offering students and educators a tool to explore concepts numerically, mathematically, and algorithmically.





















\section*{Acknowledgments}

This work is supported by the National Science Foundation Graduate Research Fellowship Program under Grant No.~DGE-1610403. Any opinions, findings, and conclusions or recommendations expressed in this material are those of the author(s) and do not necessarily reflect the views of the National Science Foundation.


\bibliographystyle{bibtex/IEEEtran}
\bibliography{bibtex/IEEEabrv,refs}

\begin{thebibliography}{1}
\providecommand{\url}[1]{#1}
\csname url@samestyle\endcsname
\providecommand{\newblock}{\relax}
\providecommand{\bibinfo}[2]{#2}
\providecommand{\BIBentrySTDinterwordspacing}{\spaceskip=0pt\relax}
\providecommand{\BIBentryALTinterwordstretchfactor}{4}
\providecommand{\BIBentryALTinterwordspacing}{\spaceskip=\fontdimen2\font plus
\BIBentryALTinterwordstretchfactor\fontdimen3\font minus
  \fontdimen4\font\relax}
\providecommand{\BIBforeignlanguage}[2]{{%
\expandafter\ifx\csname l@#1\endcsname\relax
\typeout{** WARNING: IEEEtran.bst: No hyphenation pattern has been}%
\typeout{** loaded for the language `#1'. Using the pattern for}%
\typeout{** the default language instead.}%
\else
\language=\csname l@#1\endcsname
\fi
#2}}
\providecommand{\BIBdecl}{\relax}
\BIBdecl

\bibitem{heath_lozano}
R.~W. Heath~Jr. and A.~Lozano, \emph{Foundations of {MIMO}
  Communication}.\hskip 1em plus 0.5em minus 0.4em\relax Cambridge University
  Press, 2018.

\bibitem{mfm}
I.~P. Roberts, ``{MIMO} for {MATLAB}: A toolbox for simulating {MIMO}
  communication systems in {MATLAB},'' \url{http://mimoformatlab.com}, 2021.

\bibitem{balanis}
C.~Balanis, \emph{Antenna Theory: Analysis and Design}.\hskip 1em plus 0.5em
  minus 0.4em\relax John Wiley \& Sons, 2016.

\bibitem{heath_overview_2016}
R.~W. Heath, N.~Gonz{\'a}lez-Prelcic, S.~Rangan, W.~Roh, and A.~M. Sayeed, ``An
  overview of signal processing techniques for millimeter wave {MIMO}
  systems,'' \emph{{IEEE} J. Sel. Topics Signal Process.}, vol.~10, no.~3, pp.
  436--453, Apr. 2016.

\bibitem{spherical_2005}
J.-S. Jiang and M.~A. Ingram, ``Spherical-wave model for short-range {MIMO},''
  \emph{{IEEE} Trans. Commun.}, vol.~53, no.~9, pp. 1534--1541, Sep. 2005.

\end{thebibliography}

\end{document}